\documentclass[12pt]{article}
\usepackage{a4}
\usepackage{amsfonts,amssymb,amsmath,amsthm}
\usepackage{graphicx}
\usepackage{cite}
\usepackage{slashed}
\begin{document}

\title{Quantum aspects of a noncommutative supersymmetric kink}
\author{
D.~V.~Vassilevich\thanks{On leave from V.~A.~Fock Institute of
Physics, St.~Petersburg University, Russia.
E.mail:\ {\texttt{dmitry(at)dfn.if.usp.br}}}\\
{\it Instituto de F\'isica, Universidade de S\~ao Paulo,}\\ {\it
Caixa Postal 66318, CEP 05315-970, S\~ao Paulo, S.P., Brazil}}

\maketitle
\begin{abstract}
We consider quantum corrections to a kink of noncommutative supersymmetric
$\varphi^4$ theory in $1+1$ dimensions. Despite the presence of an infinite
number of time derivatives in the action, we are able to define supercharges
and a Hamiltonian by using an unconventional canonical formalism. We calculate
the quantum energy $E$ of the kink (defined as a half-sum of the 
eigenfrequencies of fluctuations) which coincides with its' value in
corresponding commutative theory independently of the noncommutativity
parameter. The renormalization also proceeds precisely as in the
commutative case. The vacuum expectation value of the new Hamiltonian
is also calculated and appears to be consistent with the value of the
quantum energy $E$ of the kink.
\end{abstract}

\section{Introduction}
The study of quantum corrections to solitons in $1+1$ dimensions started in
1970's \cite{Dashen:1974cj,Goldstone,Rajaraman,FK}, 
and since that time a considerable
progress has been made (see \cite{Izquierdo:2006ds} for a recent review).
Noncommutative (NC) solitons \cite{Douglas:2001ba,Lechtenfeld:2006iz} 
were included
in these studies only recently \cite{Kurkcuoglu:2007rr,Konoplya:2007xt}.
The work \cite{Kurkcuoglu:2007rr} used the small $\theta$ expansion,
while the paper \cite{Konoplya:2007xt} was concentrated on moderate and large
values of the NC parameter. Both papers left many questions unanswered,
mostly related to the renormalization and to the possibility of a smooth
extension of the results to the region of large (respectively, small)
noncommutativity. Besides, in $1+1$ dimensions one deals with time-space
noncommutativity which brings an infinite number of time derivatives
into the action, so that the very definition of energy becomes
less obvious.

Another line of research considers quantum corrections to supersymmetric
solitons \cite{Rebhan:2004vu,Shifman:2007ce}. It was found 
\cite{Rebhan:1997iv}, that naive arguments leading to zero quantum corrections
to the mass of supersymmetric solitons were incorrect, and a new anomaly
(the anomaly in the central charge 
\cite{Nastase:1998sy,Graham:1998qq,Shifman:1998zy}) was discovered. 
Taking this anomaly into account restores saturation of the BPS bound
at the quantum level.

In this paper we consider quantum correction to the mass of an NC 
supersymmetric kink in $1+1$ dimensions. Our motivation is twofold.
First, it is interesting to study the interplay between supersymmetry and
noncommutativity with this particular example. Second, supersymmetry
simplifies the structure of divergences of quantum field theory and may
help to resolve some problems existing in the non-supersymmetric case.
Practically, we adapt the methods developed earlier
in \cite{Bordag:2002dg} to the NC case. Supersymmetrization of the NC
space-time is done in the most straightforward way
\cite{Chu:1999ij,Ferrara:2000mm,Terashima:2000xq} where only the bosonic
coordinates are deformed. The model we study in here is a supersymmetric
extension on the NC $\varphi^4$ model in $1+1$ dimensions.

In time-space NC theories there are well-known difficulties with 
the construction
of a canonical formalism (due to the presence of an infinite number of
time derivatives). Besides, generically there are no locally conserved currents
corresponding to global classical symmetries. Therefore, it is a priori
unclear whether one can define supercharges in such theories. However, as
we show below, this task can be successfully addressed in the framework
of an unconventional canonical formalism \cite{Vassilevich:2005fk},
so that one can introduce supercharges whose brackets give an analog
of the Hamiltonian and a central charge. The Hamiltonian has the meaning
of the energy integrated over an interval $T$ of time. For a static field
configuration it simply reads $TE$, 
where $E$ is the energy. The main reason to call these
quantities supercharges and a Hamiltonian is that with respect to the
new brackets they indeed generate global supesymmetry transformations and
the equations of motion, respectively. 

Static solutions in NC models in $(1+1)$ dimensions are not deformed, i.e.
they are the same as in corresponding commutative models. The equations of
motion for small fluctuations above such solutions are deformed, and the
fluctuations are described by wave equations with frequency-dependent 
potentials. Nevertheless, in the model we consider, bosonic and fermionic
modes are isospectral. To use all advantages of the isospectrality, we employ
the zeta-function regularization and make the spectrum discrete by introducing
boundaries in the spatial direction. (These boundaries are removed at the
end of the calculations). The quantum energy is defined as one half the sum
over the eigenfrequencies. The width of the effective potential in the wave
equations for the fluctuations with the frequency $\omega$ is 
proportional to $\theta\omega$, where $\theta$ is the NC parameter.
To keep boundaries far away from the location of the potential we have
to make the position of the boundaries frequency-dependent 
\cite{Konoplya:2007xt}. In this approach, quantum energy of the kink
is defined as the energy of a system consisting of the kink and the
boundaries minus the (Casimir) energy of the boundaries \cite{Bordag:2002dg}. 
For the renormalization, we use the heat kernel subtraction scheme which
was shown to be equivalent to the no-tadpole condition in the commutative
case \cite{Bordag:2002dg}. The divergences are removed by a renormalization
of the mass, which is precisely the same as in the commutative case. The 
renormalized
energy (mass shift of the kink) does not depend on $\theta$
and coincides with its' commutative value. 

Keeping in mind future applications to the verification of the quantum BPS
bound saturation, we also calculate quantum corrections to the new 
Hamiltonian. We find the value $TE$, where $E$ is the mass shift
of the soliton. Two apparently different definitions of the quantum
energy give consistent results. Also, the renormalization required 
for the Hamiltonian is the same mass renormalization which we described
above.

This paper is organized as follows. In the next section we introduce
a classical action and collect some preliminary information. In section
\ref{sec-can} we study the new unconventional definition of the
canonical algebra, and define supercharge, the Hamiltonian, and the
central charge. In section \ref{sec-flu} we study the spectrum of
fluctuations above the kink. Quantum corrections to the mass of the kink
are calculated in section \ref{sec-qua}, and corrections to the new
Hamiltonian are considered in section \ref{sec-Ham}. Concluding remarks are
given in section \ref{sec-con}.
\section{The classical action}\label{sec-clas}
We shall describe noncommutativity of the space-time coordinates by
the Moyal product
\begin{equation}
(f\star g) (x)=\left[ \exp \left( \frac i2 \Theta^{\mu\nu} \partial_\mu^x
\partial_\nu^y \right)f(x)g(y) \right]_{y^\mu=x^\mu},\label{Moyal}
\end{equation}
where $\Theta^{\mu\nu}$ is a constant skew-symmetric matrix which 
can be chosen as
$\Theta^{\mu\nu} = 2\theta \epsilon^{\mu\nu}$ with $\epsilon^{01}=1$.
After splitting the coordinates into time and space, $\{ x^\mu\}=\{t,x\}$,
we have the following useful formulae
\begin{equation}
f(x) \star e^{i\omega t}=e^{i\omega t} f(x+\theta \omega),\quad
e^{i\omega t}\star f(x) =e^{i\omega t} f(x-\theta \omega)\,.
\label{shift}
\end{equation}
The Moyal product is closed,
\begin{equation}
\int d^2x\, f_1\star f_2 = \int d^2x\, f_1 \cdot f_2\,, \label{closed}
\end{equation}
and has the property that
\begin{equation}
\int d^2x\, f_1\star f_2 = (-1)^{g_1g_2}
\int d^2x\, f_2 \star f_1\,, \label{cycl}
\end{equation}
where the grading $g_i=0$ if $f_i$ is bosonic, and $g_i=1$ if $f_i$ is
fermionic. To derive the properties (\ref{closed}) and (\ref{cycl}) one
has to integrate by parts in (\ref{Moyal}). In general, boundary terms
may appear. To avoid them, we assume that in the time direction all fields
are periodic with a very large period which should be sent to infinity
at the end.  In the spatial directions all fields must approach constant
values sufficiently fast. Such boundary conditions are satisfied by
static solitons and classical variations of the fields which produce the
equations of motion. A different set of boundary conditions will be used
in sec.\ \ref{sec-flu} to analyze quantum fluctuations.

The action for a supersymmetric NC $\varphi^4$ model reads
\begin{equation}
S=-\frac 12 \int_{\mathcal{M}} d^2x\, \left( (\partial_\mu \varphi)^2
+U'(\varphi )\star \bar\psi\star \psi +\bar\psi \gamma^\mu \partial_\mu \psi
-2F\star U -F^2 \right).\label{suact}
\end{equation}
Here $\varphi$ is a real scalar field, and $\psi$ is a Majorana
spinor. We take $\gamma$-matrices in the Majorana representation
\begin{equation}
\gamma^0 =-i\sigma^2
= \left( \begin{array}{cc} 0 & -1 \\ 1 & 0\end{array} \right),
\qquad
\gamma^1 =\sigma^3 =\left( \begin{array}{cc} 1&0\\0&-1 \end{array} \right).
\label{v-mgamma}
\end{equation}
In this representation the components of $\psi$ are real. $\bar\psi =
\psi^T i\gamma^0$. Components of the spinors will be marked by the 
subscripts $\pm$, so that $\psi =\left( \begin{array}{c} \psi_+ \\
\psi_- \end{array} \right)$, $\epsilon =\left( \begin{array}{c} \epsilon_+ \\
\epsilon_- \end{array} \right)$. For the $\varphi^4$ model
\begin{equation}
U(\varphi)=\sqrt {\frac {\lambda}2 } (v_0^2 - \varphi\star\varphi),\qquad
U'(\varphi)= -\sqrt{2\lambda} \varphi\,.\label{UUpr}
\end{equation}
Note, that though due to (\ref{closed}) one star can always be deleted
under an integral, it is more convenient to write all stars explicitly 
in all terms higher that second order in the fields since mixed (star
with ordinary) products are not associative. 

The supersymmetry transformations 
\begin{equation}
\delta\varphi =\bar\epsilon\psi,\qquad \delta\psi=
(\gamma^\mu\partial_\mu \varphi +F)\epsilon,\qquad
\delta F=\bar\epsilon \gamma^\mu \partial_\mu \psi .
\label{Fsusy}
\end{equation}
are linear, and, therefore, are undeformed. The invariance of (\ref{suact})
under (\ref{Fsusy}) follows from the general analysis of 
\cite{Ferrara:2000mm,Terashima:2000xq}, but can also be verified directly.

The auxiliary field $F$ may be excluded by means of its' algebraic\footnote{
This means that no derivatives acting on $F$ appear.}
equation of motion
\begin{equation}
F=-U(\varphi). \label{Feq}
\end{equation}
The action (\ref{suact}) becomes
\begin{equation}
S=-\frac 12 \int_{\mathcal{M}} d^2x\, \left( (\partial_\mu \varphi)^2
+U'(\varphi )\star \bar\psi\star \psi +\bar\psi \gamma^\mu \partial_\mu \psi
+ U\star U \right),\label{noFact}
\end{equation}
and the supersymmetry transformations read
\begin{equation}
\delta\varphi =\bar\epsilon\psi,\qquad \delta\psi=
(\gamma^\mu\partial_\mu \varphi -U)\epsilon .
\label{susy}
\end{equation}

The equations of motion corresponding to the action (\ref{noFact}) are
\begin{eqnarray}
&&\partial_\mu \partial^\mu \varphi 
+ \sqrt{\frac {\lambda}2} \, \bar \psi\star \psi 
-\frac 12 (U\star U' + U'\star U)=0,\label{beom}\\
&&\slashed{\partial} \psi + \frac 12 ( U'\star \psi + \psi \star U')=0.
\label{feom}
\end{eqnarray}
Static solutions of these equations are the same as in the commutative case.
In particular, there is the kink solution
\begin{equation}
\Phi (x)=v_0 \tanh \left( v_0 \sqrt{\frac {\lambda}2 } \, x\right).
\label{kink}
\end{equation}
This solution satisfies the Bogomolny equation
\begin{equation}
\partial_1 \Phi (x) = U(\Phi)\label{Beq}
\end{equation}
and is invariant under the supersymmetry transformations (\ref{susy}) with
$\epsilon_-=0$.  

\section{Canonical realization of the supersymmetry algebra}\label{sec-can}
We have no locally conserved supercurrent in the model
(as there is no locally conserved energy-momentum tensor
in NC theories \cite{Gerhold:2000ik}), but still
by using an unconventional canonical formalism 
for time-space noncommutative theories \cite{Vassilevich:2005fk}
we can define supercharges which generate the supersymmetry transformations
(\ref{susy}). Let us briefly outline the formalism of \cite{Vassilevich:2005fk}
(in \cite{Vassilevich:2005fk} only the bosonic case was considered, but
an extension to the presence of fermions is straightforward). The canonical
pairs are defined ignoring the time derivatives hidden in the star-product.
In our model this implies that they are precisely the same as in the
commutative case. To read off the symplectic form, let us re-write
the action (\ref{noFact}) in a ``hamiltonian'' form
\begin{eqnarray}
S&=& \int d^2x \left( -\frac i2 (\partial_0 \psi_+ \cdot \psi_+
+\partial_0\psi_- \cdot \psi_-)+\frac 12 ((\partial_0\varphi)p-
(\partial_0p)\varphi) -\mathcal{H} \right)\label{Hamact}\\
&=& \int d^2x \left( -\frac 12 (C^{-1})^{AB}\partial_0 z_A \cdot z_B
-\mathcal{H}\right) .\nonumber
\end{eqnarray}
(We use the conventions of Henneaux \cite{Henneaux:1985kr}). Here 
$\{ z_A\} \equiv \{ \varphi,p,\psi_+,\psi_-\}$. 
\begin{equation}
{\mathcal{H}}=\frac 12 \left( (\partial_1 \varphi)^2 + p^2 +
U\star U + U'\star \bar \psi \star \psi +\bar\psi \gamma^1 \partial_1\psi
\right) \label{H}
\end{equation}
does not contain explicit time derivatives (all time derivatives are
hidden in the star product).

The canonical brackets are taken between
variables at different times and are postulated to be proportional to
{\it two-dimensional} delta-functions instead of one-dimensional ones,
$\{ z_A(t,x),z_B(t',x'\}=C_{AB}\delta(t-t')\delta(x-x')$. More
explicitly,
\begin{eqnarray}
&&\{ \varphi (t,x),p(t',x') \}=\delta (x-x')\delta (t-t'),\label{canrel1}\\
&&\{ \psi_\pm (t,x),\psi_\pm (t',x') \}=-i \delta (x-x')\delta (t-t'),
\label{canrel2}
\end{eqnarray}
and $p=\partial_0 \varphi$. Usual grading rules are understood. Now we have
to extend the definition of the brackets to star-polynomials of $z_A$ and their
derivatives. Here we face a difficulty since a star product by a
delta-function is not a well defined object. However, we can define brackets
between space-time integral of polynomials. Let $F$, $G$ be two such 
integrals. Then
\begin{equation}
\{ F,G\} = \int d^2x \frac {\delta^r F}{\delta z_A(x)}\star C_{AB}
\frac {\delta^l G}{\delta z_B(x)} \,.\label{FGbr}
\end{equation}
Here $\delta^r$ and $\delta^l$ are right and left variational derivatives.
For a practical use, the formula (\ref{FGbr}) has to be understood in
the following way. One has to take all pairs of canonical variables $z_A$,
$z_B$ in $F$ and $G$ respectively, then one uses the property (\ref{cycl})
to bring $z_A$ to the rightmost position in $F$, and $z_B$ to the leftmost
position in $G$. Then one integrates by parts to 
remove all explicit derivatives
form $z_A$ and $z_A$. Then one deletes $z_A$ and $z_B$, star-multiply
the expressions obtained, contracts with $C_{AB}$ and integrates over the
space-time. The brackets defined in this way satisfy the (graded) Jacobi
identities. For bosonic theories this was demonstrated in 
\cite{Vassilevich:2005fk}, and an extension to fermions is straightforward.

By taking $F=\int f\star \hat F$, where $f$ is a smooth function,
calculating the bracket with $G$, and then varying with respect to
$f$, one can extend the definition to brackets between star-polynomials
$\hat F$ and integrated star-polynomials $G$. This trick does not work
twice. Therefore, it is not possible to define a bracket between
unintegrated polynomials, but we shall not need such an object.

In \cite{Vassilevich:2005fk} it was shown that these unconventional Poisson
brackets can be used to define first-class constraints and generate gauge
transformations in time-space NC theories (see also \cite{Vassilevich:2006uv}
for an example of practical use of these brackets). Here we shall apply
them to analyze global symmetries.

First we note, that if we define the ``Hamiltonian'' as a space-time integral
\begin{equation}
H=\int d^2x\, \mathcal{H} \label{HH}
\end{equation}
of the density (\ref{H}), then the brackets with $H$ generate the equations
of motion
\begin{equation}
\{ H,z_A\} = -\partial_0 z_A\,. \label{Hameq}
\end{equation}

A definition of the ``supercharge'' then follows by an educated guess as
a suitable generalization of corresponding commutative expression. Let us take
\begin{equation}
Q=-\int d^2x (\slashed{\partial}\varphi +U(\varphi))\star
\gamma^0 \psi\,. \label{Q}
\end{equation}
It is easy to check that this ``supercharge'' indeed generates the supesymmetry
transformations 
\begin{equation}
\{ \bar \epsilon Q,z_A \} =-\tilde\delta z_A \label{Qsusy}
\end{equation}
of the Hamiltonian action (\ref{Hamact}). On shell the transformations
$\tilde \delta$ coincide with (\ref{susy}).

We see, that the ``Hamiltonian'' and the ``supercharge'' possess
the characteristic features which we expect from a Hamiltonian and
a supercharge. Therefore, we shall sometimes omit the quotation marks in what
follows. 

The kink solution (\ref{kink}) is invariant under the $\epsilon_+$ 
transformations, which are generated by $Q_-$. The bracket of two such 
supercharges reads
\begin{equation}
\{ Q_-,Q_-\} = -2i (H-Z) \label{QQ}
\end{equation}
where\footnote{Note, that there is another total derivative term in
$\{  Q_-,Q_-\}$, namely $-i\int \partial_1 (\bar\psi \psi)$. This term
vanishes if one considers fluctuations above the kink solution
with the asymptotic conditions we discussed above. However,
such terms are important for the ``supersymmetry without
boundary conditions'' approach \cite{Belyaev:2008xk}.}
\begin{eqnarray}
&&Z=\int d^2x \partial_1 W(\varphi),\label{Z}\\
&&W(\varphi) =\sqrt{\frac {\lambda}2 } \left( v_0^2 \varphi -\frac 13 
\varphi \star \varphi \star \varphi \right) .\label{W}
\end{eqnarray}
$Z$ is a natural generalization of the central charge to the NC case.
We obtained a standard form of a central extension of the supersymmetry
algebra in a topologically non-trivial sector \cite{Witten:1978mh}, 
though the generators are given by two-dimensional integrals and the
brackets are unconventional. 

On the kink background both $H$ and $Z$ are divergent unless one
restricts the integration over $t$ to a finite interval.
Note, that the difference $H-Z$ for the kink is finite
and vanishes.
\section{Fluctuations}\label{sec-flu}
The spectrum of fluctuations is defined by the linearized equations of motion
(\ref{beom}) and (\ref{feom}). For the fermionic fluctuations we have
\begin{equation}
\left( \begin{array}{cc} \partial_1 +\frac 12 (L( U'(\Phi))+
R(U'(\Phi))) & -\partial_0 \\
\partial_0 & -\partial_1 + \frac 12 (L( U'(\Phi))+
R(U'(\Phi))) \end{array} \right)
\left( \begin{array}{c} \psi_+ \\ \psi_- \end{array} \right) =0 .
\label{linDir}
\end{equation}
Here $L$ and $R$ denote left and right Moyal multiplications respectively,
$f_1\star f_2=L(f_1)f_2=R(f_2)f_1$.
The fluctuation operator commutes with $\partial_0$. Consequently, we
can look for the solutions in the form
\begin{equation}
\psi_\pm (t,x)=e^{i\omega_f t} \psi_\pm (\omega_f,x) .
\label{v-psiom}
\end{equation}
The equation (\ref{linDir}) then yields
\begin{eqnarray}
&&i\omega_f \psi_+ (\omega_f,x)=(\partial_1 -\frac 12 ( U'(\Phi_+)+
U'(\Phi_-))
)\psi_-(\omega_f,x),
\nonumber\\
&&i\omega_f \psi_- (\omega_f,x)=(\partial_1 +\frac 12 ( U'(\Phi_+)+
U'(\Phi_-)))\psi_+(\omega_f,x),
\label{Dir2}
\end{eqnarray}
where
\begin{equation}
\Phi_{\pm}(x) \equiv \Phi (x\pm \theta\omega)\,. \label{Phipm}
\end{equation}
The property (\ref{shift}) of the Moyal product has been used.
By iterating the equations (\ref{Dir2}) one obtains
\begin{eqnarray}
&&\omega_f^2 \psi_+ (\omega_f,x)= 
- D_-(\omega_f)D_+(\omega_f) \psi_+ (\omega_f,x),
\nonumber\\
&&\omega_f^2 \psi_- (\omega_f,x)= 
- D_+(\omega_f)D_-(\omega_f) \psi_- (\omega_f,x),
\label{Dir3}
\end{eqnarray}
where
\begin{equation}
D_\pm (\omega)= 
\partial_1 \mp \sqrt{\frac{\lambda}2} (\Phi_++\Phi_-) .\label{v-Dpm}
\end{equation}

In the bosonic sector, we decompose the scalar field as
$\varphi =\Phi +\phi$. The fluctuations $\phi$ satisfy the linearized field
equation
\begin{equation}
-\partial_0^2\phi = -(\partial_1^2 +\lambda v_0^2 -
\lambda (L(\Phi^2)+R(\Phi^2)+L(\Phi)R(\Phi))\phi \,.\label{scalin1}
\end{equation}
Again, we look for the solutions in the form $\phi (\omega_b,x)=
e^{i\omega t} \phi (\omega_b,x)$. The equation (\ref{scalin1}) yields
\begin{equation}
\omega_b^2 \phi (\omega_b) = -(\partial_1^2 +\lambda v_0^2 -
\lambda (\Phi^2_++\Phi^2_-+\Phi_+\Phi_-))\phi(\omega_b). \label{scalin2}
\end{equation}
By using the Bogomolny equation (\ref{Beq}) we obtain
\begin{equation}
\omega_b^2 \phi(\omega_b) =-D_+(\omega_b)D_-(\omega_b)\phi(\omega_b).
\label{scalin3}
\end{equation}

The spectrum of the eigenfrequencies is defined by two operators,
$P_1(\omega)=-D_+(\omega)D_-(\omega)$ and 
$P_2(\omega)=-D_-(\omega)D_+(\omega)$. Due to the intertwining
relations
\begin{equation}
P_1(\omega)D_+(\omega)=D_+(\omega)P_2(\omega),\qquad
D_-(\omega)P_1(\omega)=P_2(\omega)D_-(\omega) \label{inter}
\end{equation}
these operators are isospectral up to zero modes. Indeed, these relations
imply that if $P_1\psi_1=\lambda \psi_1$, then $D_-\psi_1$ is an eigenfunction
of $P_2$ with the same eigenvalue. Also, if $P_2\psi_2=\lambda \psi_2$,
then $P_1(D_+\psi_2)=\lambda (D_+\psi_2)$. 

An explicit form of $P_1$ follows from (\ref{scalin2}). For the sake of
completeness we also present
\begin{equation}
P_2(\omega)=-(\partial_1^2 -\lambda v_0^2 -\lambda \Phi_+\Phi_-).
\label{P2}
\end{equation}

In calculations of the quantum corrections it is convenient to go from the
continuous to discrete spectrum of $P_1$ and $P_2$ by introducing boundaries 
\cite{Bordag:2002dg} in the $x$-direction. We like the boundary to interact 
with the soliton as weak as possible. Therefore, the boundary should be far
away from the place where the kink is localized. However, as we see e.g.
from eq.\ (\ref{scalin2}), the width of the effective potential is
proportional to $\theta\omega$ and becomes infinite for $\omega\to\infty$.
No boundary seems to be sufficiently far away. To overcome this difficulty,
in \cite{Konoplya:2007xt} it was suggested to make the boundary 
$\omega$-dependent, i.e. to place it to the points 
$x=\pm l(\omega)=\pm (l_0+\theta\omega)$ with a large $l_0$. 
Having a boundary, one has to impose some boundary conditions 
on the fluctuations. Particular choice of the boundary conditions
is not too important (as anyway we are going to subtract the vacuum
energy related to the boundary), but too use the full strength
of supersymmetry it is convenient to take supersymmetric boundary
conditions which respect the intertwining relations (\ref{inter})
and, therefore, preserve isospectrality of $P_1(\omega)$ and 
$P_2(\omega)$ for any $\omega$. The simplest choice is to impose the
Dirichlet boundary conditions on $\phi$ and $\psi_-$,
\begin{equation}
\phi \vert_{x=\pm l(\omega)}=
\psi_- \vert_{x=\pm l(\omega)}=0.\label{Dircon}
\end{equation}
The intertwining relations then require a Robin (generalized Neumann) boundary
condition for $\psi_+$:
\begin{equation}
D_+\psi_+ \vert_{x=\pm l(\omega)}=0.\label{Robcon}
\end{equation}
(Note that the same boundary condition on $\psi_+$ follows from
the consistency of the Dirac equation (\ref{Dir2})).

In general, the Moyal product cannot be restricted to
an interval with frequency dependent boundaries. However, for
operators commuting with the time derivatives (in particular,
for Moyal multiplications by a time-independent function)
such a restrictions can be made along the lines described in
this section.

\section{Quantum corrections to the mass}\label{sec-qua}
Here we use a generalization of the method \cite{Bordag:2002dg}
to the NC case. Namely, we first consider the kink with boundaries
with fluctuations subject to the boundary conditions (\ref{Dircon})
and (\ref{Robcon}), calculate the total quantum energy of this system
$E_{\rm k+b}$, and then subtract the vacuum energy $E_{\rm b}$
which is due to the presence of boundaries. The energy associated
with the kink is then
\begin{equation}
E_{\bf k}=E_{\rm k+b}-E_{\rm b}.\label{Ekbk}
\end{equation}


The vacuum energy for each of the systems is defined as a half-sum of
the eigenfrequencies,
\begin{equation}
E=\frac 12 \sum \omega_b -\frac 12\sum \omega_f \label{Esum}
\end{equation}
(we set $\hbar=1$). In time-space NC theories there is no standard
canonical Hamiltonian to justify this formula for the energy
(though, there is a non-standard one, see sec.\ \ref{sec-can}
and \ref{sec-Ham}). For systems with a finite number of additional time
derivatives (with fields in
stationary but non-static geometries being an example of such systems)
it was shown that this definition of the energy is equivalent to
the canonical one, and the presence of extra time derivative
(which results in modifications of the Klein-Gordon current and
corresponding scalar product) influences the results of quantum
computations through modification of the spectral density
\cite{Fursaev:2000dv,Fursaev:2001yu} (see also 
\cite{Strelchenko:2007xh,Konoplya:2007xt} for an extension of this analysis
to NC case). Adopting the same approach here looks as the most reliable
extension of the notion of vacuum energy to time-space NC theories.

Because of the presence of boundaries we deal with a
discrete spectrum of eigenfrequencies. It is convenient to use
the zeta-function regularization \cite{Dowker:1975tf,Hawking:1976ja}. 
The operator $P_1$ (resp., $P_2$) is a product of a first-order operator
and its' formal adjoint. Therefore, both $P_1$ and $P_2$ are non-negative.
In the positive spectrum, the zeta-regularized energy reads
\begin{equation}
E(s)=\frac {\mu^{2s}}2 \left( {\sum}' (\omega_b^2)^{\frac 12 -s}
- {\sum}' (\omega_f^2)^{\frac 12 -s}\right),\label{zregE}
\end{equation}
where prime tells us that the summation runs over the positive spectrum
only. (Zero frequencies do not contribute to the vacuum energy). 
The parameter $\mu$ of the dimension of the mass is introduced in order to
keep right dimensionality of the energy independently of the regularization
parameter $s$. Both sums on the right hand side of (\ref{zregE})
are convergent for ${\rm Re}\,(s)$ sufficiently large. At the end of
the calculations the result must be analytically continued to the
physical value $s=0$.

Let us first analyze $E_{\rm k+b}$. Due to the isospectrality properties
discussed above 
\begin{equation}
E(s)_{\rm k+b}=0, \label{Eskb}
\end{equation}
i.e., the regularized vacuum energy vanished identically.

Although, obviously, the vacuum energy (\ref{Eskb}) is not divergent,
there might be some finite contribution due to a finite renormalization
\footnote{This indeed happens in some models. For example, 
the whole correction to the mass of
the supersymmetric Abrikosov-Nielsen-Olesen vortex is due to a finite
renormalization of couplings \cite{Vassilevich:2003xk,Rebhan:2003bu}.}.
To define such a contribution one should fix a normalization condition
or a subtraction scheme. Here we use the   
heat-kernel subtraction scheme which is frequently
employed in the Casimir energy calculations and is discussed in detail in
\cite{Bordag:1998vs,Bordag:2001qi}. Consider a (bosonic)
system in $1+1$ dimensions
with a discrete frequency spectrum $\{ \omega_n \}$. Let 
$k_n^2=\omega_n^2-m^2$, where $m$ is the mass (or, the asymptotic value
of the potential). The regularized vacuum energy for this system admits
a representation,
\begin{equation}
\frac {\mu^2}2 \sum_n (k_n^2+m^2)^{\frac 12 -s}=
\frac {\mu^2}2 \int_0^\infty \frac{d\tau}{\tau}\,
\frac {\tau^{s-\frac 12}}{\Gamma \left(s-\frac 12\right)}\,
K(\tau)e^{-\tau m^2} \,,\label{intrep}
\end{equation}
where 
\begin{equation}
K(\tau)=\sum_n e^{-\tau k_n^2} \label{Ktau}
\end{equation}
is the corresponding heat kernel. Usually, the heat kernel admits an asymptotic
expansion\footnote{Such an expansion indeed exists for practically all case
appearing in the context of quantum field theory. A more precise and complete
information on the heat kernel expansion can be found in 
\cite{Vassilevich:2003xt} for commutative space, and in 
\cite{Vassilevich:2007fq} in the NC case. The heat kernel for 
frequency-dependent problems was analyzed in 
\cite{Fursaev:2001yu,Fursaev:2001fm,Fursaev:2002vi}.}
\begin{equation}
K(\tau)\simeq \sum_{p>0} a_p \tau^{p-1}\label{asym}
\end{equation}
as $\tau\to +0$. For $s=0$ contributions to (\ref{intrep})
from the terms with $p=0,1,2$
are divergent at the lower limit. We define the divergent part of
the vacuum energy as
\begin{eqnarray}
&&E^{\rm div}\equiv \frac {\mu^2}2 \int_0^\infty \frac{d\tau}{\tau}\,
\frac {\tau^{s-\frac 12}}{\Gamma \left(s-\frac 12\right)}\,
\sum_{n=0}^2 a_n \tau^{n-1} e^{-\tau m^2}\nonumber\\
&&\qquad = \frac {\mu^2}{2\Gamma \left(s-\frac 12\right)}
\left\{ a_0 \Gamma (s-1)m^{2-2s}+a_1 \Gamma  \left(s-\frac 12\right)
m^{1-2s}\right.\nonumber\\
&&\qquad\qquad \left. +a_2 \Gamma (s)m^{-2s}\right\} . \label{Ediv}
\end{eqnarray}
The renormalized energy is then
\begin{equation}
E^{\rm ren}=[E(s) - E^{\rm div}(s)]_{s=0} \label{defEren}
\end{equation}
This subtraction scheme has two important advantages. First, in the case
of commutative scalar theories in $1+1$ dimensions it is equivalent
\cite{Bordag:2002dg}
to the ``no tadpole'' normalization condition which is commonly used
to calculate the mass shift of two-dimensional solitons. Second, 
this scheme can easily be extended to the NC case.

Let us return to $E_{\rm k+b}$. Due to (\ref{Eskb}) the heat kernel
is also identically zero, as well as all heat kernel coefficients
and $E^{\rm div}_{\rm k+b}(s)$. We conclude, that
\begin{equation}
E^{\rm ren}_{\rm k+b}=0.\label{Ekbren}
\end{equation}

Next we have to study the vacuum energy $E_{\rm b}$ due to the presence
of boundaries. Far away from the kink, the excitations are
free bosonic and fermionic modes with the mass $m=v_0\sqrt{2\lambda}$
which is defined by asymptotic values of the potential in (\ref{scalin2})
and (\ref{P2}). In the bosonic sector, the boundary conditions are Dirichlet.
In the fermionic sector, one mode satisfies the Dirichlet conditions as
well, another one satisfies the Robin boundary conditions (each of the
modes carries one half of a degree of freedom)\footnote{It is important
that, as in the commutative case \cite{Bordag:2002dg}, we use for the
fermions an asymptotic form  of the squared Dirac equation (\ref{Dir3}).
One cannot substitute asymptotic values of the fields in the Dirac
equation (\ref{linDir}) itself and then extend it smoothly to the whole
space $[-l,l]$.}.

Let us study the Robin sector first. For large $l_0$, (we remind
that $l(\omega)=l_0+\theta\omega$) the condition
(\ref{Robcon}) yields
\begin{equation}
(\partial_x + S_1)\psi\vert_{x=-l(\omega)}=0,\qquad
(-\partial_x + S_2)\psi\vert_{x=l(\omega)}=0,
\label{Nbc}
\end{equation}
where
\begin{equation}
S_1=S_2=v_0 \sqrt{2\lambda}\equiv S.\label{S1S2}
\end{equation}
There are no bound states ($\omega^2<m^2$) for these boundary conditions.
The spectrum of oscillating modes, $\psi = A\sin (kx) + B\cos (kx)$,
$k=\sqrt{\omega^2 -m^2}$ is given by solutions of the equation
\cite{Bordag:2002dg}
\begin{equation}
0=f(\alpha_1,\alpha_2;k)\equiv
\sin (2kl(\omega)+\alpha_1+\alpha_2) \label{fk}
\end{equation}
with
\begin{equation}
\alpha_{1,2}=-\arctan (k/S_{1,2}) \equiv \alpha \,.\label{a12}
\end{equation}
It is easy to see, that the spectrum in the Dirichlet sector is defined
by the equation
\begin{equation}
0=f(0,0;k)\,.\label{fDir}
\end{equation}

Next we represent the vacuum energy as a contour integral 
\cite{Bordag:1994jz,Bordag:2002dg}. The function $\partial_k \ln f(k)$
has poles with unit residues at the points where $f(k)=0$. Therefore,
we can write 
\begin{equation}
E_{\rm b}(s)=
-\frac {\mu^{2s}}4 \oint \frac{dk}{2\pi i} (k^2+m^2)^{\frac 12 -s} 
\frac \partial{\partial k} (\ln f(\alpha,\alpha;k) -\ln f(0,0;k)),
\label{AEC1}
\end{equation}
where the contour goes anticlockwise around the positive real semiaxis.
Along the upper part of the contour we approximate 
$\sin (2(kl(\omega)+\alpha))$ by $-(1/2i)\exp (-2i(kl(\omega)+\alpha))$
since the term $\exp (2i(kl(\omega)+\alpha))$ vanishes as $l_0\to\infty$.
Along the lower part we keep $(1/2i)\exp (2i(kl(\omega)+\alpha))$.
Then,
\begin{equation}
E_{\rm b}(s)=-\mu^{2s} \int_0^\infty \frac{dk}{2\pi }
(k^2+m^2)^{\frac 12 -s}\, \frac {\partial\alpha}{\partial k}\,. \label{Esbfin}
\end{equation}
We see, that all contributions containing $l(\omega)$ are cancelled.
Therefore, the regularized boundary energy is given by precisely 
the same expression  as in the
commutative case (cf.\ \cite{Bordag:2002dg}). 
Without any further calculations we can read off
the renormalized value
\begin{equation}
E_{\rm b}^{\rm ren}=\sqrt{\lambda/2}\, \frac {v_0}{\pi}\label{Ebren}
\end{equation}
from \cite{Bordag:2002dg}. Consequently, the renormalized vacuum energy
of the kink
\begin{equation}
E_{\rm k}^{\rm ren}=E_{\rm b+k}^{\rm ren}-E_{\rm b}^{\rm ren}
=-\sqrt{\lambda/2}\, \frac {v_0}{\pi} \label{Ekren}
\end{equation}
does not depend on $\theta$ and coincides with its' value in the
commutative theory.
\section{Vacuum expectation value of the new canonical Hamiltonian}
\label{sec-Ham}
There is little doubt in the correctness of the definition of the
 vacuum energy used in the previous section. However, keeping in
 mind the applications to saturation of the BPS bound one should also
 calculate corrections to the new Hamiltonian (\ref{HH}) which
 participates in the supersymmetry algebra.

To calculate vacuum expectation value of the Hamiltonian
(\ref{HH}) we need the propagators for small fluctuations over the
kink background. Let us start in the bosonic sector. Consider
eigenfunctions of the operator $P_1(\omega)$,
\begin{equation}
P_1(\omega) \tilde\phi_{\omega,\lambda_\omega}(x)=\lambda_\omega^2
\tilde\phi_{\omega,\lambda_\omega}(x)\,,\label{phomlom}
\end{equation}
and normalize them according to the condition
\begin{equation}
\int dx\, \tilde\phi_{\omega,\lambda_\omega}^*(x)
\tilde\phi_{\omega,\lambda'_\omega}(x)=\delta_{\lambda_\omega,\lambda'_\omega}
\,.\label{normphi}
\end{equation}
We assumed, that there is a boundary in the $x$-direction, so that the 
spectrum is discrete. The functions $\tilde \phi_{\omega,\lambda_\omega}$
are defined initially on the interval $[-l(\omega),l(\omega)]$ but can be
extended to the whole $\mathbb{R}$ as $\tilde \phi_{\omega,\lambda_\omega}=0$
for $|x|>l(\omega)$. The operator $P_1(\omega)$ acts by its analytic formula
inside the interval and is extended as multiplication by $\lambda_\omega^2$
outside the interval and on the boundary. (Of course, as long as $l(\omega)$
is finite the functions $\tilde\phi_{\omega,\lambda_\omega}$ cannot be used
to expand an arbitrary function on $\mathbb{R}$).
The integration in (\ref{normphi})
can run over $\mathbb{R}$, but the dual formula
\begin{equation}
\sum_\lambda \tilde \phi^*_{\omega,\lambda_\omega}(x)
\tilde \phi_{\omega,\lambda_\omega}(x')=\delta(x-x')
\label{nck35}
\end{equation}
is valid only if both $x$ and $x'$ belong to  $[-l(\omega),l(\omega)]$.
Otherwise, the right hand side is zero.

The functions
\begin{equation}
\phi_{\omega,\lambda_\omega}(x^\mu)=e^{-i\omega t}
\tilde \phi_{\omega,\lambda_\omega}(x) \label{phfull}
\end{equation}
are the eigenfunctions of the full kinetic operator acting
on fluctuations (restricted to an interval) with eigenvalues
$-\omega^2+\lambda_\omega^2$. The propagator can then be
constructed in the standard way as
\begin{equation}
G(x^\mu,{x^\mu}')=\frac 1{2\pi}\int d\omega \sum_{\lambda_\omega}
\frac{ \phi_{\omega,\lambda_\omega}(x^\mu)
\phi_{\omega,\lambda_\omega}^*({x^\mu}')}{-\omega^2+\lambda_\omega^2
-i\varepsilon},\label{GF}
\end{equation}
but the relation $P_1G(x^\mu,{x^\mu}')=\delta (x^\mu{x^\mu}')$
is true only if both $x^1$ and ${x^1}'$ belong to the intersection of
the intervals  $[-l(\omega),l(\omega)]$, i.e., to  $[-l_0,l_0]$.
For $l_0\to \infty$ one recovers the Feynman propagator.
Then,
\begin{equation}
\langle \phi (x^\mu) \phi (y^\nu)\rangle = -iG(y^\nu,x^\mu).
\label{vevpp}
\end{equation}
With the help of this equation one calculates the one-loop
vacuum expectation of the bosonic part of the Hamiltonian
\begin{eqnarray}
&&\langle H \rangle_B =-\frac i2  
\int d^2x (-\partial_0^2-\partial_1^2 
+\lambda v_0^2 -\lambda (L(\Phi^2)+R(\Phi^2)+L(\Phi)R(\Phi))_x
\nonumber\\
&&\qquad\qquad\qquad\qquad\qquad\qquad \times
\left. G(x^\mu,y^\nu)\right|_{y^1=x^1,\ y^0=x^0+\sigma}.
\label{vevHb}
\end{eqnarray}
where we introduced a time-splitting regularization
with the parameter $\sigma$. The operator acting on $G$ should be understood
as $-\partial_0^2+P_1$. The action of $P_1$ on 
$\tilde\phi_{\omega,\lambda_\omega}$
is already defined above. 
It is easy to see, that the integrand does not depend on $x^0$. In
order to remove the corresponding divergence we restrict the integration
over $x^0$ to $[0,T]$ with some finite $T$. We have,
\begin{eqnarray}
&&\langle H \rangle_B =-\frac {iT}2 \int \frac {d\omega}{2\pi}\int dx^1
\sum_{\lambda_\omega} \frac {\omega^2+\lambda_\omega^2}{-\omega^2+
\lambda_\omega^2 -i\varepsilon}
\tilde \phi_{\omega,\lambda_\omega}(x^1)
 {\tilde \phi_{\omega,\lambda_\omega}}^* (x^1)\, e^{i\omega\sigma}\nonumber\\
&&\qquad = -\frac {iT}2 \int \frac {d\omega}{2\pi}
\sum_{\lambda_\omega} \frac {\omega^2+\lambda_\omega^2}{-\omega^2+
\lambda_\omega^2 -i\varepsilon}\, e^{i\omega\sigma}\label{vevH2}
\end{eqnarray}
Let $\sigma<0$. The integration contour can be closed in the lower
complex half-plane.
For each value of $\omega$ there is a discrete set of eigenvalues
$\{ \lambda^j_\omega\}$. Let $\omega_j$ be positive solutions
of the equation $\omega_j=\lambda^j_{\omega_j}$ (there could be
multiple solutions of this equation for each $j$, but we do not consider
such case for simplicity). Then,
\begin{equation}
\langle H \rangle_B = \frac T2 \sum_j \omega_j \left( 1 - 
\frac{d\lambda^j_\omega}{d\omega}\vert_{\omega=\omega_j}\right)^{-1}.
\label{vevH3}
\end{equation}
For $\sigma >0$ the result is the same.

This formula admits a rather simple interpretation. The factor $T$ 
appears since our Hamiltonian has the meaning of energy integrated over
the time. The expression under the sum is an energy of an excitation
with the frequency $\omega_j$. In the commutative limit the derivative
in the bracket vanishes, so that each excitation contributes $\frac 12 \omega$.
In the NC case, a correction factor appears. The presence of this factor means
that the contribution of an individual mode to $\langle H \rangle$
differs from that to $E$. As we shall see below, due to the supersymmetry
this difference does not affect the final result when contributions of
all modes, bosonic and fermionic, are taken into account.

For contribution of the fermionic fluctuations one obtains 
similarly\footnote{The only subtlety is the way to extend the eigenfunctions
satisfying Robin boundary conditions outside the interval $[-l(\omega),
l(\omega)]$. This should be done again by setting these function to zero.
Possible discontinuities at the boundary do not play a role. In this way
we preserve the isospectrality of $P_1$ and $P_2$.}
\begin{equation}
\langle H \rangle_F = -\frac T2 \sum_j \omega_j \left( 1 - 
\frac{d\lambda^j_\omega}{d\omega}\vert_{\omega=\omega_j}\right)^{-1},
\label{vevH4}
\end{equation}
where, as expected, the overall sign is different from (\ref{vevH3}).
$\omega_j$ now denote the fermionic frequencies.

Due to the isospectrality of bosonic and fermionic fluctuations on
a background of the kink in the presence of boundaries
\begin{equation}
\langle H \rangle_F^{\rm b+k} +\langle H \rangle_B^{\rm b+k}=0 .
\label{Hbk}
\end{equation}
(It is understood that these quantities must be regularized by 
replacing $\omega$ with $\omega^{1-2s}$. The calculations proceed
precisely as in the previous section.)

Let us now calculate the boundary contribution $\langle H \rangle^{\rm b}$
to the vacuum expectation value of the Hamiltonian. An effective free field
theory which must be used to calculate boundary contributions was described
in the previous section. The boundary conditions are given by (\ref{Dircon})
and (\ref{Robcon}), and the spectrum of $\lambda_\omega^j$ is defined by
the solutions of the equation $f(\omega | \lambda_\omega)=0$, where
\begin{equation}
f(\omega | \lambda)=\sin (2(k(\lambda)l(\omega)+\alpha(k(\lambda)))),
\qquad k(\lambda)=\sqrt{\lambda^2-m^2}, \label{fol}
\end{equation}
$\alpha(k)=0$ for Dirichlet conditions, $\alpha(k)=-\arctan (S/k)$ for
Robin ones. The quantity
\begin{equation}
h(s)=\sum_j  \omega_j^{1-2s} \left( 1 - 
\frac{d\lambda^j_\omega}{d\omega}\vert_{\omega=\omega_j}\right)^{-1},
\end{equation}
which is a zeta-regularized expression for the right hand sides of
(\ref{vevH3}) and (\ref{vevH4}), can be represented as a contour
integral
\begin{equation}
h(s)=\frac 1{2\pi i} \oint d\omega\, \omega^{1-2s} \left( 1 - 
\frac{d\lambda_\omega}{d\omega}\vert_{\omega=\lambda_\omega}\right)^{-1}
\partial_\omega (\ln f(\omega \vert \omega ) ),\label{hs}
\end{equation}
where the contour encircles $[m,\infty[$. One can write
\begin{equation}
\partial_\omega f(\omega | \omega) =[\partial_\omega f(\omega |\lambda)
+\partial_\lambda f(\omega |\lambda)]_{\lambda =\omega}.\label{209}
\end{equation}
On the other hand, the condition $f(\omega |\lambda_\omega)=0$ defines the
dependence of $\lambda_\omega$ on $\omega$. By differentiating this condition,
one gets
\begin{equation}
0=\partial_\omega f(\omega |\lambda_\omega)=
[\partial_\omega f(\omega|\lambda)]_{\lambda =\lambda_\omega} +
[\partial_\lambda f(\omega|\lambda)]_{\lambda =\lambda_\omega} 
\frac{d\lambda_\omega}{d\omega}.\label{215}
\end{equation}
By using (\ref{209}) and (\ref{215}) we rewrite (\ref{hs}) 
as\footnote{This equation can be also obtained in a different way. As follows
from the analysis of \cite{Fursaev:2001yu,Strelchenko:2007xh}, the factor
$(1-(d\lambda/d\omega))^{-1}$ is the difference between the spectral density
of eigenfrequencies $\omega_j$ and the spectral density of the eigenvalues
$\lambda$ for a given $\omega$ taken at $\lambda=\omega$. The integral
(\ref{hs2}) is simply a sum over the eigenfrequencies with the latter
density.}
\begin{equation}
h(s)=\frac 1{2\pi i} \oint d\omega\, \omega^{1-2s}
[\partial_\lambda \ln f(\omega |\lambda )]_{\lambda =\omega}.
\label{hs2}
\end{equation}

By using the identities which we have just derived one can represent 
the zeta-regularized boundary contribution to the v.e.v.\ of $H$ in
the form
\begin{equation}
\langle H \rangle^{\rm b}(s)=
\frac {T\mu^{2s}}4 \, \frac 1{2\pi i}\oint d\omega\, \omega^{1-2s}
[\partial_\lambda (\ln f_D(\omega |\lambda )-
\ln f_R(\omega |\lambda ))]_{\lambda =\omega}\,,\label{Hbs}
\end{equation}
where $f_{D,R}$ correspond to Dirichlet and Robin boundary conditions
respectively (cf.\ eq.\ (\ref{fol}) and the line below). As in the previous
section, on the upper part of the contour we approximate 
$\sin (2(kl(\omega)+\alpha)$ by $-(1/2i)\exp (-2i(kl(\omega)+\alpha))$,
and by $(1/2i)\exp (2i(kl(\omega)+\alpha))$ on the lower part.  Then the 
terms with $l(\omega)$ cancel, and we arrive at the expression
\begin{equation}
\langle H \rangle^{\rm b}(s)=-T\mu^{2s} \int_m^\infty \frac{d\omega}{2\pi}\,
\omega^{1-2s}[\partial_\lambda \alpha (k(\lambda))]_{\lambda =\omega}.
\label{235}
\end{equation}
Next we observe that $[\partial_\lambda \alpha (k(\lambda))]_{\lambda =\omega}
=\partial_\omega \alpha (k(\omega))$ with $k(\omega)=\sqrt{\omega^2-m^2}$
and change the integration variable to $k$.
\begin{equation}
\langle H \rangle^{\rm b}(s)=-T\mu^{2s}
\int_0^\infty \frac{dk}{2\pi}\, (k^2+m^2)^{\frac 12 -s} \partial_k \alpha (k),
\label{239}
\end{equation}
or,
\begin{equation}
\langle H \rangle^{\rm b}(s)=TE^{\rm b}(s). \label{240}
\end{equation}
In the heat kernel subtraction scheme 
$\langle H \rangle^{\rm div}(s)=TE^{\rm div}(s)$, so that for the renormalized
values we also have the relation
\begin{equation}
\langle H \rangle^{\rm b}_{\rm ren}=TE^{\rm b}_{\rm ren}.
\label{306}
\end{equation}
Taking into account (\ref{Ekbren}) and (\ref{Hbk}), we conclude that
\begin{equation}
\langle H \rangle^{\rm k}_{\rm ren}=
TE^{\rm k}_{\rm ren}= -T\sqrt{\lambda/2}\, \frac{v_0}{\pi}.\label{Hkren}
\end{equation}
This is a very natural result. It tells us that the interpretation
 of the new canonical Hamiltonian as the energy integrated over a
 time interval remains valid also at the one-loop level.

\section{Conclusions}\label{sec-con}
In this work we studied quantum corrections to the mass of the
 kink of supersymmetric NC $\varphi^4$. Contrary to the
 nonsupersymmetric case \cite{Konoplya:2007xt}, the counterterm
 required to remove the divergences is precisely the same as in the
 commutative theory. The strategy of calculations of the one-loop
 corrections was taken from \cite{Bordag:2002dg}. We introduced
 boundaries, so that the spectrum of the fluctuations becomes
 discrete. Because of the nonlocality of NC theories, the position
 of the boundary depends on the frequency of each fluctuation. For
 the system of the kink and the boundaries, we used the
 isospectrality of bosonic an fermionic fluctuations which follows
 from supersymmetry. The total energy of this system vanishes. Then
 we subtracted the contribution from the boundaries, which was
 calculated in a relatively simple effective theory. The heat
 kernel subtraction scheme (which is equivalent to the "no-tadpole"
 normalization condition in two-dimensional commutative models)
 gave a value of the mass correction which did not depend on the
 NC parameter and coincided with the commutative value.
 
 By making use of an unconventional canonical formalism we were
 able to define supercharges (despite the presence of an infinite
 number of time derivatives and the absence of locally conserved
 currents), and to show that the new brackets of these supercharges
 give an analog of the Hamiltonian and an analog of the central
 charge. (Note, that the supercharges do generate the supersymmetry
 transformations, and the Hamiltonian does generate the equations
 of motion, provided the new canonical brackets are used). This
 Hamiltonian can be interpreted as the energy integrated over an
 interval $T$ of the time. The one-loop vacuum expectation value of
 this Hamiltonian appears to be the quantum correction to the mass
 of the kink times $T$, i.e., the picture remains consistent after
 turning on the quantum effects. Although we have two different
 definitions of quantum corrections to the energy (one through a
 sum over the eigenfrequencies, and the other through the
 Hamiltonian of the unconventional canonical formalism), both
 definitions give essentially equivalent results.
 
 In a future publication we are going to calculate quantum
 corrections to the central charge. This will allow to check
 whether the quantum BPS bound remains saturated in NC theories. It
 would also be interesting to consider quantum corrections to
 solitons in higher dimensional NC theories.

\section*{Acknowledgments}
This work was supported in part by FAPESP and CNPq.


\end{document}